\documentclass[aps,prl,preprint,groupedaddress,showpacs]{revtex4}
\usepackage[dvips]{graphicx}

\begin{document}
\preprint{KUNS-1750}
\title{Wobbling Motion in Atomic Nuclei with Positive-Gamma Shapes}

\author{Masayuki Matsuzaki}
\email[]{matsuza@fukuoka-edu.ac.jp}
\affiliation{Department of Physics, Fukuoka University of Education, 
             Munakata, Fukuoka 811-4192, Japan}

\author{Yoshifumi R. Shimizu}
\email[]{yrsh2scp@mbox.nc.kyushu-u.ac.jp}
\affiliation{Department of Physics, Graduate School of Sciences, Kyushu University,
             Fukuoka 812-8581, Japan}

\author{Kenichi Matsuyanagi}
\email[]{ken@ruby.scphys.kyoto-u.ac.jp}
\affiliation{Department of Physics, Graduate School of Science, Kyoto University,
             Kyoto 606-8502, Japan}


\begin{abstract}
The three moments of inertia associated with the wobbling mode built on the 
superdeformed states in $^{163}$Lu are investigated 
by means of the cranked shell model plus random phase 
approximation to the configuration with an aligned quasiparticle. 
The result indicates that it is crucial to take into account the direct 
contribution to the moments of inertia from the aligned quasiparticle 
so as to realize $\mathcal{J}_x > \mathcal{J}_y$ in positive-gamma shapes. 
Quenching of the pairing gap cooperates with the alignment effect.
The peculiarity of the recently observed $^{163}$Lu data is discussed 
by calculating not only the electromagnetic properties but also the 
excitation spectra. 
\end{abstract}

\maketitle

 Rotation is one of the specific collective motions in finite many-body 
systems. Most of the nuclear rotational spectra can be understood 
as the outcome of one-dimensional (1D) rotations of axially symmetric nuclei. 
Two representative models --- the moment of inertia of the irrotational 
fluid, $\mathcal{J}^\mathrm{irr}$, and that of the rigid rotor, 
$\mathcal{J}^\mathrm{rig}$, both specified by an appropriate 
axially-symmetric deformation parameter $\beta$ 
--- could not reproduce the experimental ones, $\mathcal{J}^\mathrm{exp}$; 
$\mathcal{J}^\mathrm{irr}<\mathcal{J}^\mathrm{exp}<\mathcal{J}^\mathrm{rig}$. 
From a microscopic viewpoint, the moment of inertia can be 
calculated as the response of the many-body system to an externally forced 
rotation --- the cranking model~\cite{ing}. This reproduces 
$\mathcal{J}^\mathrm{exp}$ well by taking into account the pairing correlation. 
Triaxial nuclei can rotate about their three principal axes and the corresponding 
three moments of inertia depend on their shapes in general. In spite of a lot of 
theoretical studies, their shape (in particular the triaxiality parameter 
$\gamma$) dependence has not been understood well because of the lack of decisive 
experimental data. Recently some evidences of three-dimensional (3D) rotations 
have been observed, such as the shears bands and the so-called chiral-twin 
bands~\cite{frau}. In addition to these fully 3D motions, from the general 
argument of symmetry breaking, there must be a low-lying 
collective mode associated with the symmetry reduction from a 1D 
rotating axially symmetric mean field to a 3D rotating triaxial one. 
This is called the wobbling mode. Notice that the collective mode 
associated with the ``phase transition" from an axially symmetric to a triaxial 
mean field in the non-rotating case is the well known gamma vibration. Therefore 
the wobbling mode can be said to be produced by an interplay of triaxiality and 
rotation. The wobbling mode is described as a small amplitude fluctuation of the 
rotational axis away from the principal axis with the largest moment of inertia. 
Bohr and Mottelson first discussed this mode~\cite{bm}. Mikhailov and 
Janssen~\cite{mj} and Marshalek~\cite{ma} described this mode in terms of 
the random phase approximation (RPA) in the rotating frame. 
In these works it was shown that at $\gamma=0$ 
this mode turns into the odd-spin members of the gamma-vibrational band while 
at $\gamma=60^\circ$ or $-120^\circ$ it becomes the precession mode built 
on top of the high-$K$ isomeric states~\cite{an}. Here we note that, according 
to the direction of the rotational axis relative to the three principal axes 
of the shape, $\gamma$ runs from $-120^\circ$ to $60^\circ$.

 Recently electromagnetic (EM) properties of the second triaxial 
superdeformed (TSD2) band in $^{163}$Lu were reported and concluded that 
the TSD2 is a wobbling band excited on the previously known yrast TSD1 
band, on the basis of comparisons to a particle-rotor model (PRM) 
calculation~\cite{lu1,lu2}. 
In conventional PRM calculations an irrotational moment of inertia,
\begin{equation}
\mathcal{J}_k^\mathrm{irr}=\frac{4}{3}\mathcal{J}_0
\sin^2{(\gamma+\frac{2}{3}\pi k)} ,
\label{eq1}
\end{equation}
where $k=$ 1 -- 3 denote the $x$, $y$ and $z$ principal axes, is assumed. 
The magnitude $\mathcal{J}_0$ is treated as an adjustable parameter although it can 
be identified as $\mathcal{J}_0=3B_2\beta^2$, where $B_2$ is the inertia parameter 
in the Bohr Hamiltonian~\cite{bo}. 
This reduces to $\mathcal{J}^\mathrm{irr}$ in the first paragraph by substituting 
$\gamma=0$ and $k=1$, and satisfies such a required property that collective 
rotations about the symmetry axes are forbidden. Since $\mathcal{J}_y^\mathrm{irr}$ 
is largest for $0<\gamma<60^\circ$ and the main rotation occurs about the axis of 
the largest inertia, the PRM with $\mathcal{J}_k^\mathrm{irr}$ can not describe 
the positive-gamma rotation, that is, the rotation about the shortest axis 
($x$ axis). Then in Refs.~\cite{lu1,lu2} the so-called gamma-reversed moment of 
inertia~\cite{hm}, $\mathcal{J}_k^\mathrm{rev}$, 
defined by inverting the sign of $\gamma$ in Eq.(\ref{eq1}), was adopted. 
Although this reproduced the measured EM properties well, 
this does not satisfy the required property mentioned above and its physical 
implications are not very clear. In this Letter, therefore, we study the moments 
of inertia associated with the wobbling motion excited on the 
positive-gamma states by means of the cranked shell model plus RPA. 
This framework does not divide the system into a valence particle and a rotor, 
and therefore can calculate the three moments of inertia of the whole system 
microscopically. We believe that this is the first step toward understanding 
the fully 3D nuclear rotations. 

 We have developed a computer code for the RPA to excitation modes built on 
configurations with arbitrary number of aligned quasiparticles (QPs). 
In this Letter we present the results for the 4 -- 6QP configurations in Gd 
isotopes and the 1QP one in $^{163}$Lu. In particular, this is the first RPA 
calculation for the rotating odd-$A$ configurations, to our knowledge. 
Note that this approach is different from the conventional particle-vibration 
coupling calculations where the RPA itself is performed for the even-even 
``core" configurations. Since the details of the formulation have already been 
given in Refs.~\cite{sm,smm}, here we describe only the outline. The QP states were 
obtained by diagonalizing the cranked triaxial Nilsson plus BCS Hamiltonian at each 
rotational frequency $\omega_\mathrm{rot}$ with adjusting chemical 
potentials to give correct average particle numbers. 
The doubly-stretched $\mathbf{l}^2$ and $\mathbf{l}\cdot\mathbf{s}$ potentials were 
adopted, and their strengths were taken from Ref.~\cite{br}. 
The RPA calculation was performed with adopting the pairing plus doubly-stretched 
$Q\cdot Q$ interaction. The existence of aligned QPs is taken into account by 
exchanging the definition of the QP creation and annihilation operators in 
an appropriate manner. Actual calculations 
were done in five major shells ($N^\mathrm{(osc)}_\mathrm{n}=$ 3 -- 7 and 
$N^\mathrm{(osc)}_\mathrm{p}=$ 2 -- 6) by using the dispersion equation~\cite{ma}, 
\begin{equation}
(\hbar\omega)^2=(\hbar\omega_\mathrm{rot})^2
\frac{(\mathcal{J}_x-\mathcal{J}_y^\mathrm{(eff)}(\omega))
      (\mathcal{J}_x-\mathcal{J}_z^\mathrm{(eff)}(\omega))}
     {\mathcal{J}_y^\mathrm{(eff)}(\omega)\mathcal{J}_z^\mathrm{(eff)}(\omega)} ,
\label{eq2}
\end{equation}
obtained by decoupling the Nambu-Goldstone mode analytically assuming 
$\gamma\neq0$. 
This equation is independent of the strengths of the interaction. Not only the 
collective wobbling mode ($\omega=\omega_\mathrm{wob}$) but also many 
non-collective modes are obtained from this equation. The effective inertia 
$\mathcal{J}_{y,z}^\mathrm{(eff)}(\omega) 
= J_{y,z}^\mathrm{(PA)}(\omega)/\Omega_{y,z}(\omega)$, defined in the 
principal-axis frame (their concrete expressions were given in Ref.~\cite{smm}), 
depend on the eigenmode while the kinematical 
$\mathcal{J}_x=\langle J_x\rangle/\omega_\mathrm{rot}$, where the expectation 
value is taken with respect to the whole system, is common to all modes. 
It should be noted that Eq.(\ref{eq2}) coincides with the original expression 
for $\omega_\mathrm{wob}$~\cite{bm} 
if $\mathcal{J}_x$ and $\mathcal{J}_{y,z}^\mathrm{(eff)}(\omega)$ are replaced 
with constant moments of inertia.

\begin{figure}
  \includegraphics[width=13cm]{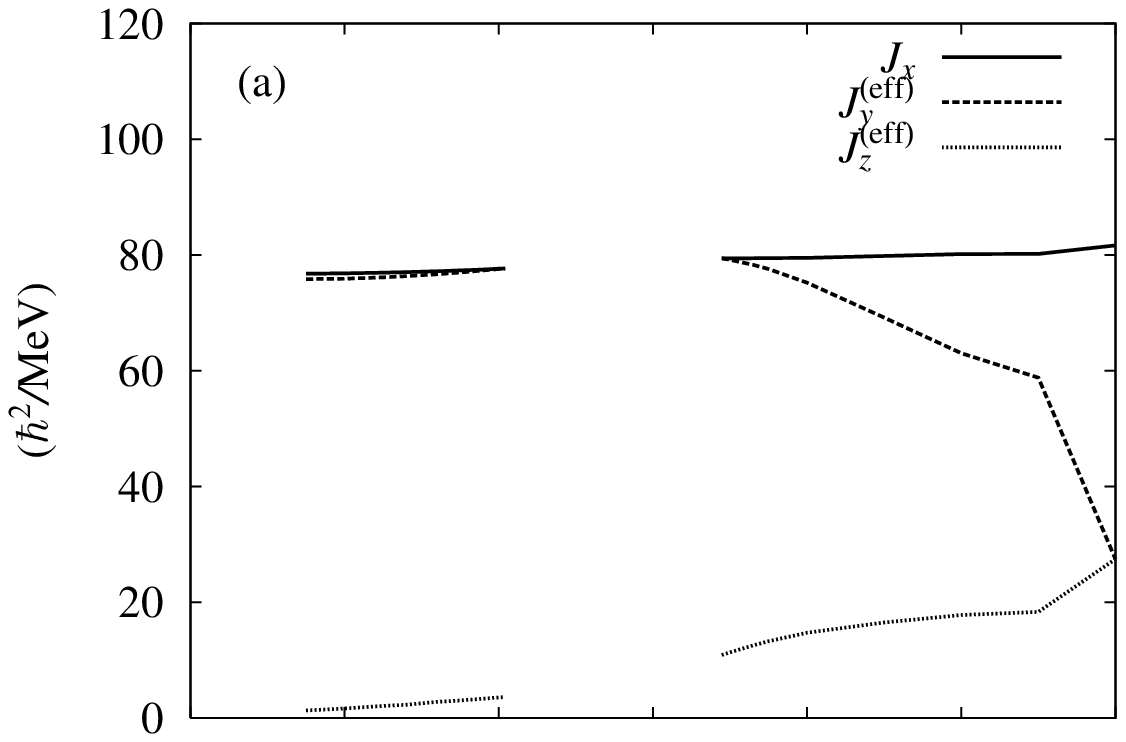}
  \includegraphics[width=13cm]{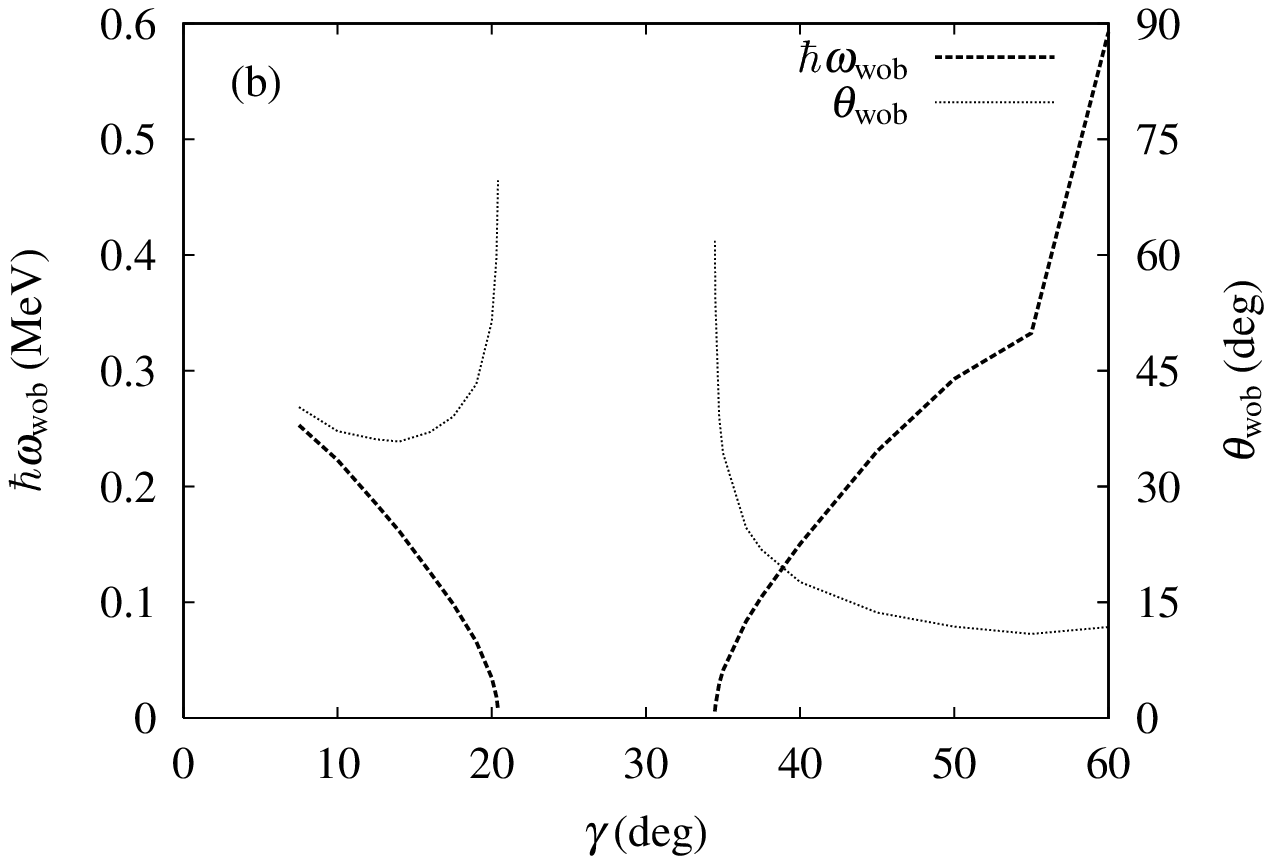}
 \caption{(a) Moments of inertia and (b) wobbling frequency (left scale) 
and wobbling angle (right) in the five 
quasiparticle state in $^{147}$Gd calculated as functions of 
$\gamma$ at $\hbar\omega_\mathrm{rot}$ = 0.3 MeV. The dip around 
$\gamma=55^\circ$ stems from a weak fragmentation of collectivity. 
Note that the present method of calculation does not 
apply to $\gamma\simeq0$. \label{fig1}}
\end{figure}

 In the following, we present some numerical results. 
Here the parameters $\epsilon_2$ (alternative to $\beta$), $\gamma$, 
$\Delta_\mathrm{n}$ and $\Delta_\mathrm{p}$ were chosen so as to reproduce 
available experimental data, and kept constant as functions of 
$\omega_\mathrm{rot}$. We have confirmed that qualitative 
features of the result are robust and the details of the parameter dependence 
will be given in a separate publication~\cite{msm_next}. 
It is non-trivial to obtain the wobbling solution in the RPA for positive-gamma 
nuclei and the QP alignment is indispensable for its appearance. 
In order to show this, we first discuss a theoretical calculation for 
a precession mode which might be built on top of the $I^\pi=49/2^+$ 
isomeric state in $^{147}$Gd, where the whole angular momentum is 
built up by the alignment of the five QPs, 
$[(\pi h_{11/2})^2(\nu h_{9/2},f_{7/2})^2]_{18^+}$ in $^{146}$Gd plus 
$[\nu i_{13/2}]_{13/2^+}$, so that a $\gamma=60^\circ$ 
shape (axially symmetric about the $x$ axis) is realized. 
This state is obtained by cranking with $\hbar\omega_\mathrm{rot}=$ 0.3 MeV. 
We chose $\epsilon_2=$ 0.19 and $\Delta_\mathrm{n}=\Delta_\mathrm{p}$ = 0.6 MeV, 
and reproduced the observed static quadrupole moment and 
$g$-factor~\cite{gd1,gd2}.  In order to see the behavior of the three moments of 
inertia, we calculated the wobbling mode by changing the parameter 
$\gamma$ from $60^\circ$. The result is presented in 
Fig.\ref{fig1}(a). Although at a first glance their $\gamma$ dependence 
resembles that of the rigid rotor, 
\begin{equation}
\mathcal{J}_k^\mathrm{rig}=\frac{16\pi}{15}B_2
\left(1-\sqrt{\frac{5}{4\pi}}\beta\cos{(\gamma+\frac{2}{3}\pi k)}\right) ,
\end{equation}
the physical contents of $\mathcal{J}_x$ changes with $\gamma$; the fraction of the 
collective contribution decreases as $\gamma$ increases and reaches 0 at 
$\gamma=60^\circ$. Accordingly it can be conjectured that the $\gamma$ dependence 
of the ``rotor" contribution is approximately irrotational and the QP contribution 
is superimposed on top of the former by aligning its angular momentum 
to the $x$ axis. 
Our previous calculation~\cite{mm,smm} for a negative-gamma nucleus, $^{182}$Os, 
also supports this and consequently it is thought that the wobbling 
mode can appear relatively easily in superfluid negative-gamma nuclei. 
To see if this conjecture is meaningful, 
starting from $^{146}$Gd we add the $i_{13/2}$ quasineutrons sequentially. 
The result shows that $\mathcal{J}_x$ increases as the number of aligned QPs 
increases. Since the increase of $\mathcal{J}_{y,z}^\mathrm{(eff)}$ is rather 
moderate, the increase of $\mathcal{J}_x$ leads to that of the wobbling 
frequency $\omega_\mathrm{wob}$. Thus, the exchange from 
$\mathcal{J}_x<\mathcal{J}_y$ in $\mathcal{J}_k^\mathrm{irr}$ to 
$\mathcal{J}_x>\mathcal{J}_y$ in $\mathcal{J}_k^\mathrm{rev}$ may be related 
qualitatively to the increase of 
$\mathcal{J}_x$  stemming from the alignment which is not accounted for in the PRM, 
considering the fact that the alignment of particle states leads to $\gamma>0$.

 At $\gamma\sim30^\circ$ where $\mathcal{J}_y$ reaches its maximum as in the 
irrotational model, we could not obtain a wobbling solution. In Fig.\ref{fig1}(b), 
$\hbar\omega_\mathrm{wob}$ and the wobbling angle, 
\begin{equation}
\theta_\mathrm{wob}=\tan^{-1}
{\frac{\sqrt{\vert J_y^\mathrm{(PA)}(\omega_\mathrm{wob})\vert^2
            +\vert J_z^\mathrm{(PA)}(\omega_\mathrm{wob})\vert^2}}
      {\langle J_x\rangle}} ,
\end{equation}
are graphed. This shows that $\omega_\mathrm{wob}$ becomes imaginary and 
$\theta_\mathrm{wob}$ blows up in this region. Comparing Fig.\ref{fig1}(a) and (b), 
it may be inferred that the wobbling motion excited on a mean field rotating about 
the $x$ axis becomes unstable at $\gamma\sim30^\circ$ due to 
$\mathcal{J}_x<\mathcal{J}_y^\mathrm{(eff)}$, 
and that a tilted-axis rotation would be realized. 
Putting this unstable region in-between, the solution in the larger-$\gamma$ side 
is like a precession of an axially symmetric body about the $x$ axis, 
whereas that in the smaller-$\gamma$ side is like a gamma vibration around 
an axially symmetric shape about the $z$ axis.

\begin{figure}
  \includegraphics[width=13cm]{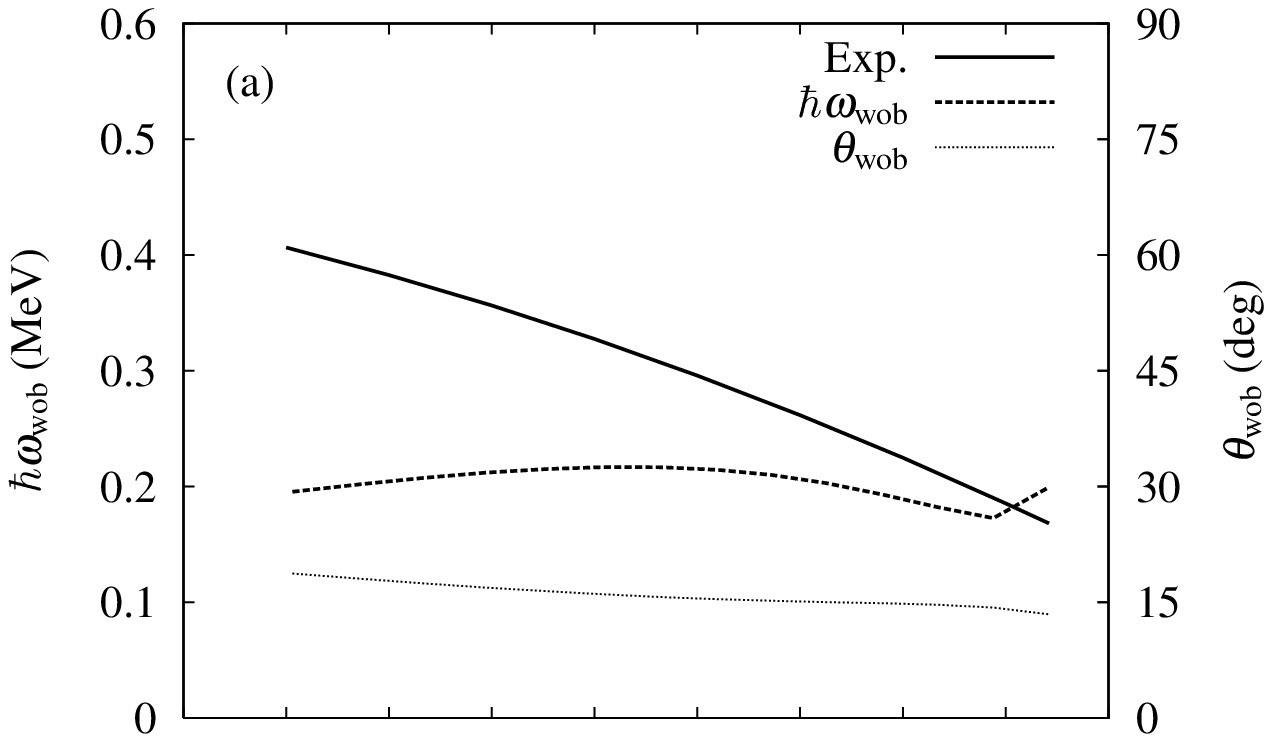}
  \includegraphics[width=13cm]{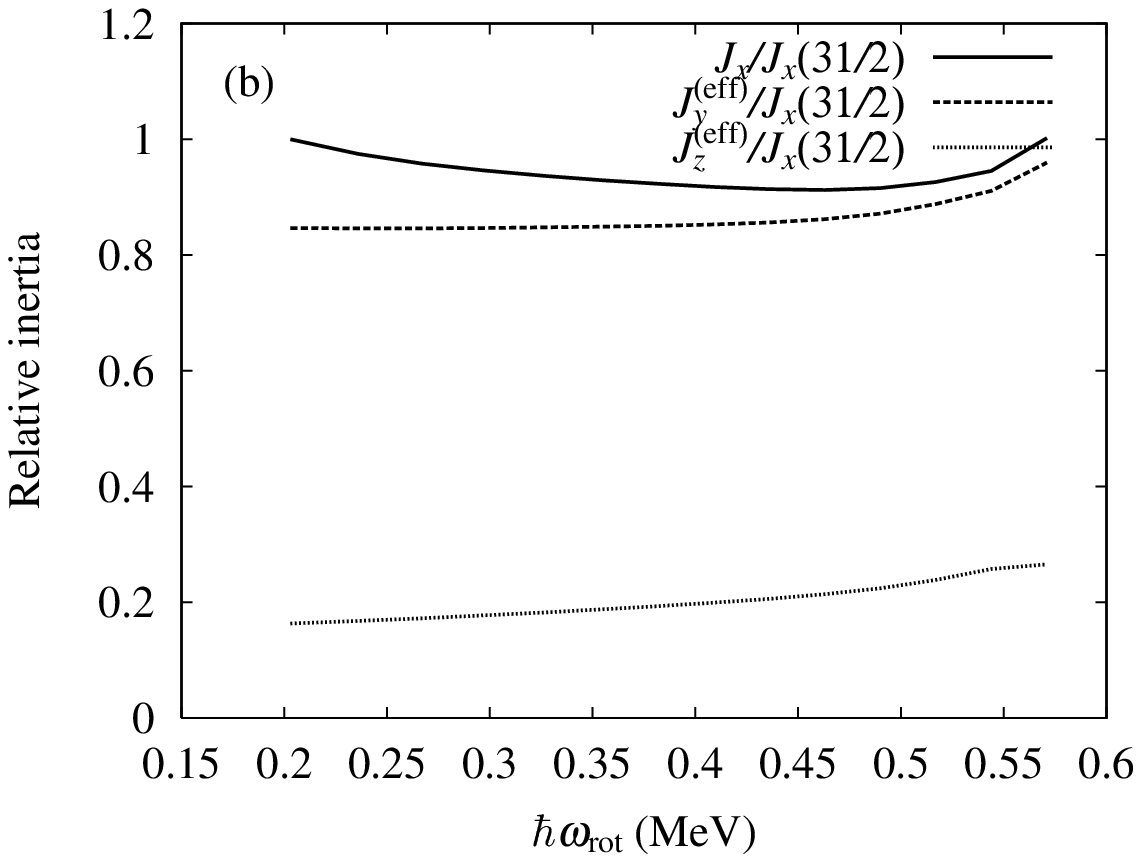}
 \caption{(a) Wobbling frequency (left scale) and wobbling 
angle (right) and (b) moments of inertia in the TSD2 band in $^{163}$Lu as 
functions of $\hbar\omega_\mathrm{rot}$. Here the latter were given by normalized 
to $\mathcal{J}_x(31/2)=99.2 \hbar^2/$MeV. The proton $BC$ crossing occurs at 
$\hbar\omega_\mathrm{rot}\agt$ 0.55 MeV in the calculation. 
Experimental values were calculated from the energy levels 
in Refs.~\cite{lu1,lu2}. \label{fig2}}
\end{figure}

 Now we turn to the TSD bands in $^{163}$Lu. We chose $\epsilon_2=$ 0.43, 
$\gamma=20^\circ$ and $\Delta_\mathrm{n}=\Delta_\mathrm{p}$ = 0.3 MeV, and obtained 
transition quadrupole moments $Q_\mathrm{t}=$ 10.9 -- 11.3 $eb$ for 
$\hbar\omega_\mathrm{rot}=$ 0.20 -- 0.57 MeV in accordance with the data, 
$Q_\mathrm{t}=10.7\pm0.7~eb$~\cite{lu0}. 
We have obtained for the first time (aside from the theoretical simulation above) 
the wobbling solution in the RPA for positive-gamma nuclei. 
Here it should be stressed that the inclusion of the five major shells and the 
alignment effect of the proton $i_{13/2}$ quasiparticle is essential for obtaining 
this result. In Fig.\ref{fig2}(a) the measured excitation energy of the TSD2 band 
relative to that of the TSD1 and the calculated $\hbar\omega_\mathrm{wob}$ are 
shown. The most peculiar point in the experimental data is that 
$\omega_\mathrm{wob}$ decreases as a function of $\omega_\mathrm{rot}$.  
If $\omega_\mathrm{rot}$-independent moments of inertia such as the irrotational 
ones are adopted, $\omega_\mathrm{wob}$ increases linearly 
with $\omega_\mathrm{rot}$, see the comment below Eq.(\ref{eq2}).  
The wobbling frequency is sensitive to the difference among the three moments of 
inertia, and the ratios $\mathcal{J}_y^\mathrm{(eff)}/\mathcal{J}_x$ and 
$\mathcal{J}_z^\mathrm{(eff)}/\mathcal{J}_x$ actually determine 
$\omega_\mathrm{wob}$. For example, the gamma-reversed moments of inertia 
give $\mathcal{J}_y^\mathrm{rev}/\mathcal{J}_x^\mathrm{rev}=0.43$ and 
$\mathcal{J}_z^\mathrm{rev}/\mathcal{J}_x^\mathrm{rev}=0.12$ for $\gamma=20^\circ$ 
leading to $\omega_\mathrm{wob} \simeq$ 3 $\omega_\mathrm{rot}$, which is quite 
different from the experimental data. In contrast, as shown in Fig.\ref{fig2}(b), 
the three moments of inertia calculated microscopically depend on 
$\omega_\mathrm{rot}$ even when the shape parameters are fixed, 
and the resultant $\omega_\mathrm{wob}$ can either increase or decrease in general. 
In the present case of $^{163}$Lu in Fig.\ref{fig2}, 
$\mathcal{J}_x-\mathcal{J}_y^\mathrm{(eff)}$ mainly determines the 
$\omega_\mathrm{rot}$ dependence.
Its decrease is a consequence of that of $\mathcal{J}_x$; 
the partial contribution to $\mathcal{J}_x$ from the proton $i_{13/2}$, 
$i_x/\omega_\mathrm{rot}$, decreases as $\omega_\mathrm{rot}$ increases 
since this orbital is already fully aligned and therefore the aligned angular 
momentum $i_x$ is approximately constant. 
Thus, our result for $\omega_\mathrm{wob}$ stays 
almost constant against $\omega_\mathrm{rot}$, and even decreases slightly 
at higer frequencies approaching the experimentally observed one. 
This clearly shows that microscopic calculation of the three moments of inertia 
is crucial to understand the $\omega_\mathrm{rot}$ dependence of 
$\omega_\mathrm{wob}$ in $^{163}$Lu. 
Let us compare this result with that for $^{147}$Gd above. In $^{147}$Gd, 
$\mathcal{J}_y^\mathrm{(eff)}/\mathcal{J}_x\simeq 1$, 
$\mathcal{J}_z^\mathrm{(eff)}\sim 0$ 
and $\vert Q_1^{(-)}/ Q_2^{(-)}\vert\ll1$ at $\gamma\alt 20^\circ$. 
The last quantity measures the rotational $K$-mixing. 
This indicates that this solution is essentially similar to 
the gamma vibration in an axially symmetric nucleus as mentioned above. 
In contrast, the result that $\mathcal{J}_y^\mathrm{(eff)}/\mathcal{J}_x= 0.90$, 
$\mathcal{J}_z^\mathrm{(eff)}/\mathcal{J}_x= 0.19$ and 
$\vert Q_1^{(-)}/ Q_2^{(-)}\vert$~= 0.78 for $^{163}$Lu at 
$\hbar\omega_\mathrm{rot}=$ 0.3 MeV, for example, indicates that this solution is 
more like a wobbling motion of a triaxial body. 
The wobbling angle shown in Fig.\ref{fig2}(a) is 19$^\circ$ -- 13$^\circ$ for the 
calculated range. It is evident that the present small-amplitude approximation 
holds better at high spins. We confirmed that this wobbling solution 
disappeared as $\gamma$ decreased. Another feature distinct from the gamma 
vibration is that the present solution exists even at 
$\Delta_\mathrm{n}=\Delta_\mathrm{p}$ = 0, whereas it is well known that the 
pairing field is indispensable for the existence of low-lying shape vibrations. 
This is related to such a tendency that the 
moments of inertia approach the rigid ones, $\mathcal{J}_x > \mathcal{J}_y$ 
for $\gamma>0$, as the pairing gap decreases even without aligned QPs. 

\begin{figure}
  \includegraphics[width=13cm]{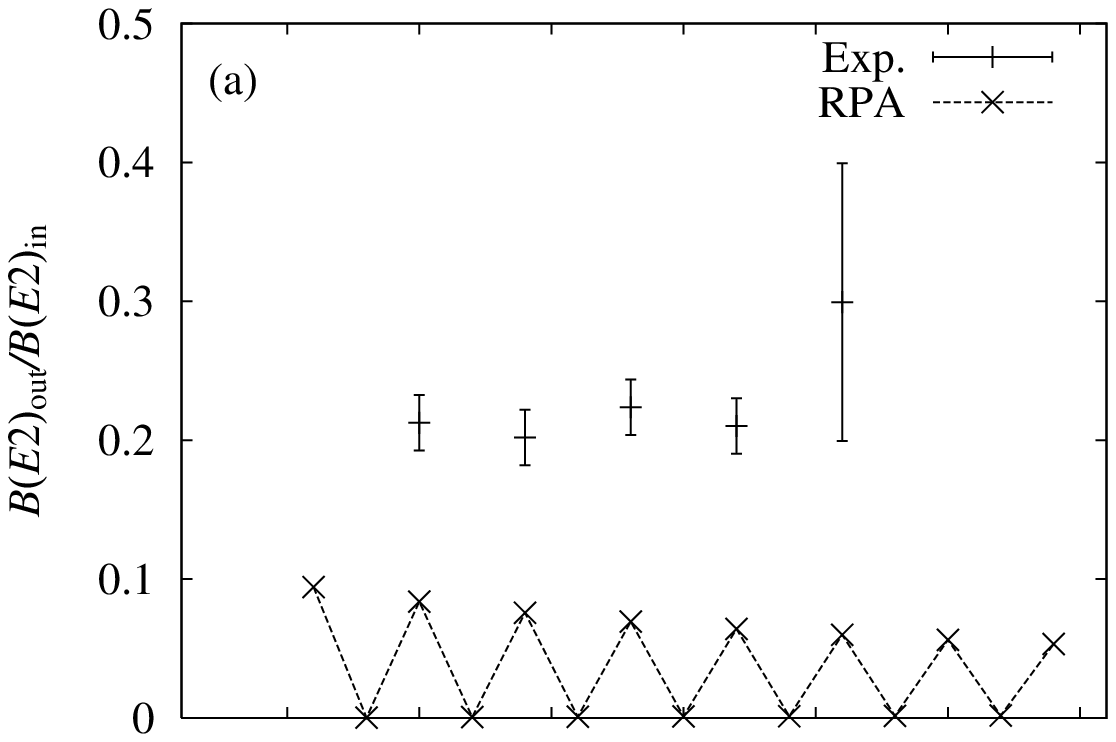}
  \includegraphics[width=13cm]{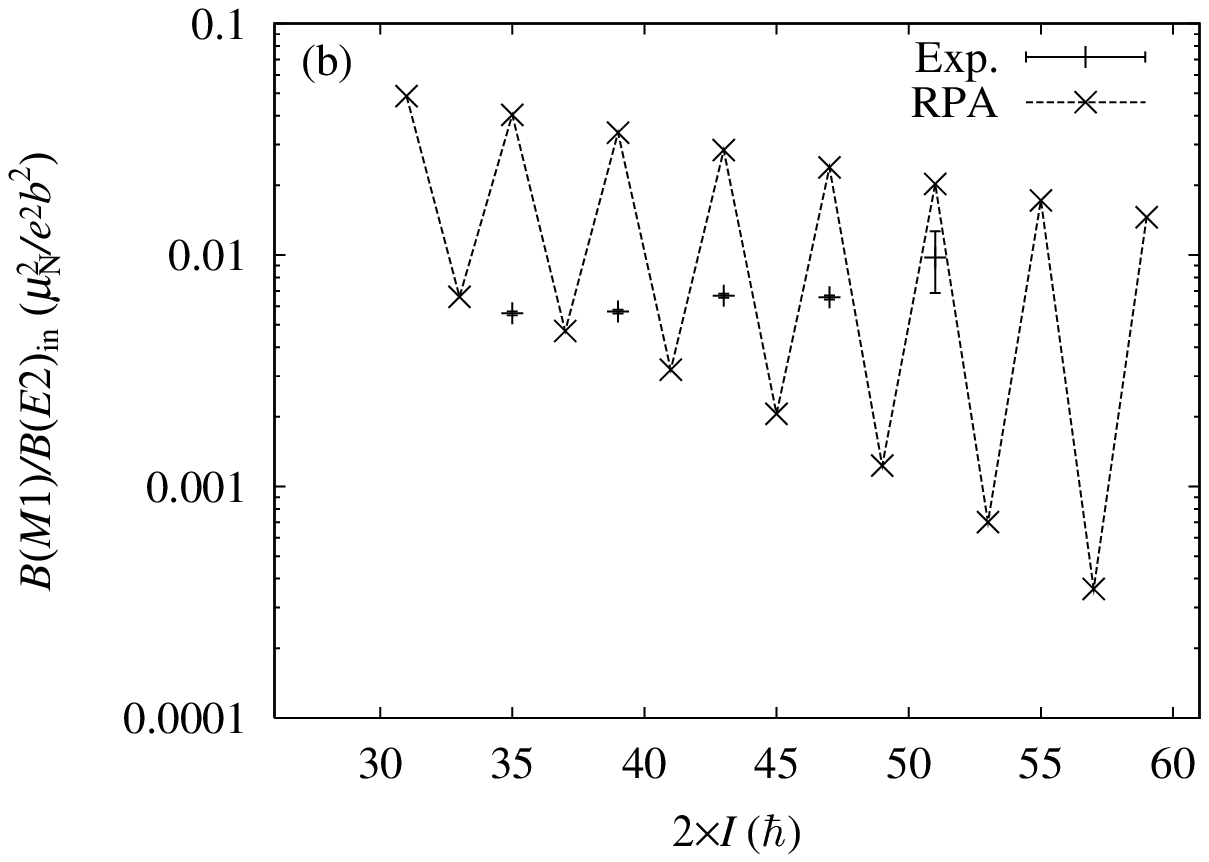}
 \caption{Interband transition rates for 
$I$ (TSD2) $\rightarrow$ $I\pm1$ (TSD1) transitions as functions of 
$2\times$ spin $I$, (a) $E2$ and (b) $M1$. They are divided by 
the in-band $E2(I \rightarrow I-2)$ transition rates. Experimental values 
were taken from Ref.~\cite{lu2}. 
Noting that the states $I+1$ (TSD1) are slightly higher in energy than 
$I$ (TSD2) at $I>51/2$ and 
$B(T_\lambda;I\rightarrow I+1)\simeq B(T_\lambda;I+1\rightarrow I)$ 
at high spins, we plotted those for $I\rightarrow I+1$ at the places with the 
abscissae $I+1$ in order to show clearly their characteristic staggering 
behavior. \label{fig3}}
\end{figure}

 A significant point of the data in Refs.~\cite{lu1,lu2} is that the interband EM 
transition rates connecting the states $I$ (TSD2) to $I-1$ (TSD1) were precisely 
measured. 
In Fig.\ref{fig3}, we compare our numerical results with the measured ones in a 
form similar to those in Refs.~\cite{lu1,lu2}. 
Calculated values for $I$ (TSD2) $\rightleftharpoons$ $I+1$ (TSD1) are also 
included in order to show the staggering behavior characteristic to this kind 
of transitions~\cite{smm}. Figure~\ref{fig3}(a) presents the 
relative $B(E2)$. The data indicate huge collectivity of the interband $B(E2)$, 
such as 170 Weisskopf unit. Although the present RPA solution is 
extremely collective, $\vert c_{n=\mathrm{wob}}\vert\simeq$ 0.9 in the sum rule 
(Eq.(4.30) in Ref.~\cite{smm}), in comparison to usual low-lying vibrations, 
the calculation accounts for 1/2 -- 1/3 of the measured strength. 
Figure~\ref{fig3}(b) graphs $B(M1)/B(E2)_\mathrm{in}$. The smallness of $B(M1)$ 
also reflects collectivity, that is, the coherence with respect to the $E2$ 
operator, indirectly. After confirming the insensitivity to 
$g_s^\mathrm{(eff)}$, we adopted 0.6 $g_s^\mathrm{(free)}$ conforming 
to Ref.~\cite{lu2} and calculated $B(M1)$. The result is similar to that of the 
PRM. We confirmed that the sign of the $E2/M1$ mixing ratios was correct.

 To summarize, 
we have performed, for the first time, the RPA calculation in the rotating 
frame to the triaxial superdeformed odd-$A$ nucleus $^{163}$Lu and discussed the 
physical conditions for the appearance of the wobbling solution in the RPA. 
We have confirmed that the proton $i_{13/2}$ alignment is 
indispensable for the appearance of the wobbling mode in this nucleus. 
The appearance of the wobbling mode requires 
$\mathcal{J}_x>\mathcal{J}_y^\mathrm{(eff)}
\left(\neq\mathcal{J}_z^\mathrm{(eff)}\right)$, but the moments of inertia 
of the even-even core exhibit irrotational-like gamma dependence and therefore can 
not fulfill this condition for positive-gamma shapes. Consequently the 
alignment effect which increases $\mathcal{J}_x$ is necessary. Quenching of 
the pairing correlation also cooperates with the alignment effect for making the 
gamma dependence rigid-like. 

\begin{acknowledgments}
We acknowledge useful discussions with I. Hamamoto and H. Madokoro. 
This work was supported in part by the Grant-in-Aid for scientific research 
from the Japan Ministry of Education, Science and Culture 
(No. 13640281 and 12640281).
\end{acknowledgments}

\end{document}